\documentstyle[psfig,epsf,conf-X]{article}

\begin{document} 
\small
\heading{%
%
MACS: The evolution and properties of massive clusters of galaxies
}
\par\medskip\noindent
\author{%
Harald Ebeling$^{1}$, Alastair Edge$^{2}$, J. Patrick Henry$^{1}$
}
\address{%
Institute for Astronomy, 2680 Woodlawn Drive, Honolulu, HI 96822, USA
}
\address{%
Department of Physics, University of Durham, Durham, DH1 3LE, UK\\*[-5mm]
}
%

\begin{abstract}
We present first results from the MAssive Cluster Survey (MACS), a
new large-scale X-ray survey designed to find and characterize 
very massive, distant clusters of galaxies. Based on X-ray detections
in the ROSAT All-Sky Survey, MACS aims to compile a sample of more
than 200 X-ray luminous clusters at $z\ge 0.3$, about 50 times the
number of EMSS clusters in the same redshift and luminosity range.
The MACS sample is uniquely suited to investigate cluster evolution
at redshifts and luminosities poorly sampled by all existing surveys.

At the time of writing the MACS sample comprises 41 clusters with
measured redshifts of $0.3\le z\le 0.56$ and X-ray luminosities in
excess of $6\times 10^{44}$ erg s$^{-1}$ (0.1--2.4 keV). This early
sample is thus already twice as large as the high-$z$, high-luminosity
subsets of the EMSS, BCS, and REFLEX cluster samples taken
together. An additional 85 MACS clusters with photometric redshifts
$z\ge 0.3$ are scheduled for spectroscopic observation.

From a preliminary analysis of a statistically well defined subsample
of the 25 X-ray brightest MACS clusters we conclude tentatively that
negative evolution is not significant at the highest X-ray
luminosities ($L_{\rm X} > 1\times 10^{45}$ erg s$^{-1}$) out to
redshifts of $z\sim 0.4$. Our findings thus extend the no-evolution
result obtained by many serendipitous ROSAT cluster surveys at lower
luminosities.\\*[-1mm]

\end{abstract}
\section{Introduction}
In a bottom-up scenario of structure formation massive galaxy clusters
form from rare, extreme overdensities in the primordial density
fluctuation field. At what redshift clusters of a given mass collapse
and virialize depends sensitively on the chosen structure formation
theory.  The comoving number density of clusters as a function of
redshift (and, ideally, also of mass) is thus an important statistic
that is well suited to constrain cosmological and physical parameters
of structure formation models. The tightest constraints can be
obtained from observations of very distant, very massive clusters
which are the rarest in all cosmological models.

Since clusters are bright X-ray sources, wide angle X-ray surveys are
an excellent way of compiling sizeable cluster samples out to
cosmological redshifts ($z\sim 1$). Several such samples have been
compiled (and/or published) in the past decade; an overview of the
solid angles and flux limits of these surveys is presented in
Figure~1. Two kinds of surveys can be distinguished: serendipitous
cluster surveys (Bright SHARC, CfA 160 deg$^2$ survey, EMSS, RDCS,
SHARC-S, WARPS) and contiguous area surveys (BCS, BCS-E, NEP, RASS-BS,
REFLEX). The former surveys use data from pointed X-ray observations,
whereas the latter are all based on the ROSAT All-Sky Survey
(RASS). With the exception of the NEP survey \cite{nep}, all
contiguous cluster surveys cover close to, or more than, 10,000 square
degrees but are limited to the X-ray brightest clusters. This
fundamental difference in depth and sky coverage has important
consequences. As shown in Fig.~1, the NEP survey as well as all
serendipitous cluster surveys (with the possible exception of the
EMSS) cover too small a solid angle to detect a significant number of
X-ray luminous clusters.  The RASS large-area surveys, on the other
hand, are capable of finding these rarest systems, but are too shallow
to detect them in large numbers at $z>0.3$.

We are thus in the unfortunate situation that the cosmologically most
important systems, the massive, distant clusters, are poorly sampled
by all existing X-ray cluster surveys. At low to moderate X-ray
luminosities where the number statistics have greatly improved most of
the above surveys find little, if any, evolution out to redshifts of
$z\sim 0.8$ \cite{warps,rdcs,cfa,borgani}. At higher luminosities,
however, the EMSS and CfA cluster surveys find evidence of strong
negative evolution already at $z>0.3$. It is worth bearing in mind
though that the latter results are based on very small samples or,
in fact, non-detections.

\begin{figure}
\vspace*{-5cm}
\mbox{\epsfxsize=9cm \hspace*{-4mm} \epsffile{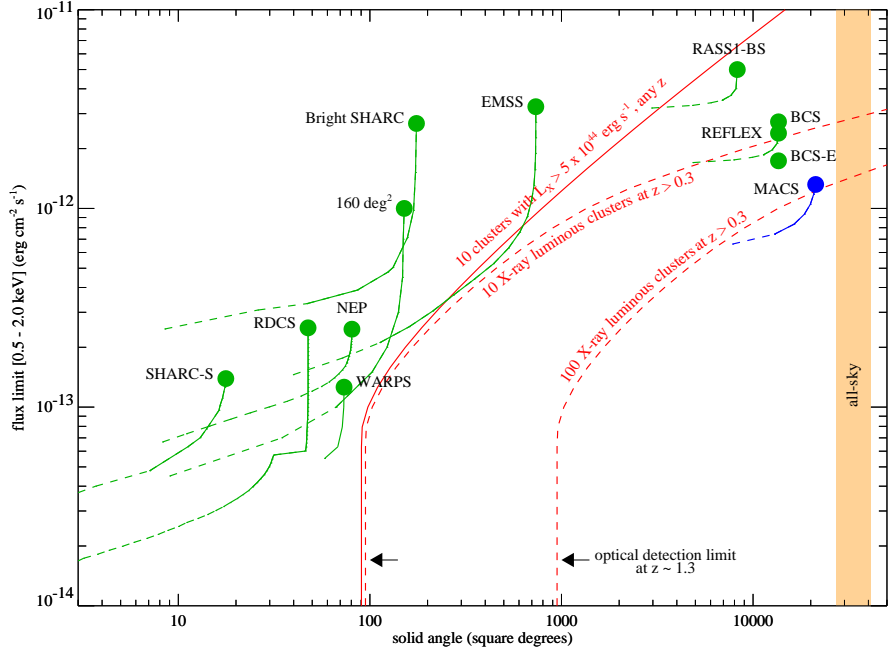}}
\caption[]{\small The selection functions of all major X-ray cluster surveys
of the past decade. Also shown is the solid angle required at a given
flux limit to (statistically) detect 10 (or 100) X-ray luminous
cluster at any redshift (or at $z>0.3$). Note how, of all previous
surveys, only the EMSS, BCS, and REFLEX projects are just sensitive
enough to detect a small number of distant, X-ray luminous systems.}
\end{figure}

\section{MACS: a new cluster survey}

MACS (MAssive Cluster Survey) was designed to find the population of
(possibly) strongly evolving clusters, i.e., the most X-ray luminous
systems at $z>0.3$. By doing so, MACS will re-measure the rate of
evolution and test the results obtained by the EMSS and CfA cluster
surveys. Unless negative evolution is very rapid indeed, MACS will
find a sizeable number of these systems and thus provide us with
targets for in-depth studies of the physical mechanisms driving
cluster evolution and structure formation.

As indicated in Fig.~1, MACS aims to achieve these goals by combining
the largest solid angle of any RASS cluster survey with the lowest
possible X-ray flux limit.  Drawing from the list of 18,000 X-ray
sources listed in the RASS Bright Source Catalog (BSC) MACS applies
the following selection criteria: $|b|\ge 20^{\circ}$, $-40^{\circ}
\le \delta \le 80^{\circ}$ (to ensure observability from Mauna Kea;
the resulting solid angle is 22,735 deg$^2$), X-ray hardness ratio
greater than a limiting ($n_{\rm H}$ dependent) value derived from
the BCS sample \cite{bcs}, and $f_{\rm X,12} \ge 1$ where $f_{\rm
X,12}$ is the detect cell flux in the 0.1--2.4 keV band in units of
$10^{-12}$ erg cm$^{-2}$ s$^{-1}$.

Cluster candidates are tentatively identified from Digitized Sky
Survey images, and are confirmed or discarded by R band imaging
observations with the University of Hawai`i's 2.2m telescope. This
process has, so far, resulted in the identification of more than 700
clusters of galaxies at all redshifts; Fig.~2 shows the redshift
distribution of the 602 systems with spectroscopic redshifts. As a
by-product, MACS has thus already delivered the by far largest X-ray
selected cluster catalogue to emerge from the RASS to date. Our
spectroscopic follow-up observations focus exclusively on the most
distant of these clusters. Redshifts are estimated from the imaging
data, and spectroscopic observations with the UH2.2m and Keck 10m
telescopes are performed of all systems with $z_{\rm{est}}>0.2$.

\begin{figure}
\mbox{\epsfxsize=10cm \hspace*{5mm} \epsffile{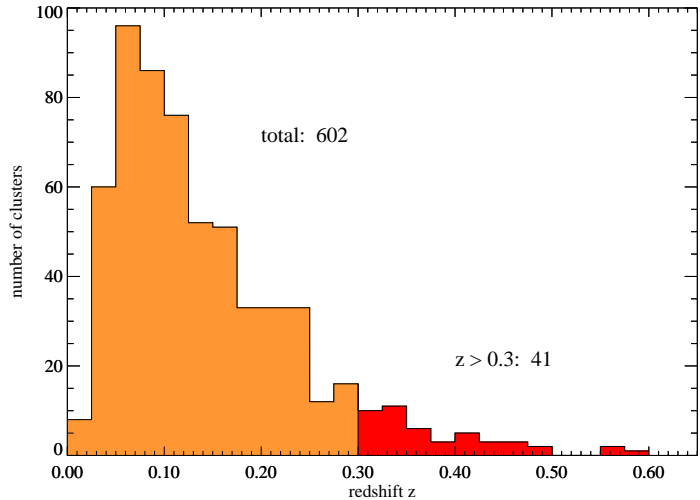}}
\caption[]{\small The redshift distribution of the 602 clusters
identified in the MACS project to date. The 41 clusters at $z>0.3$
that form the preliminary MACS sample are highlighted. Note that
all 602 clusters have spectroscopic redshifts. 
}
\end{figure}

A prediction for the size of the final MACS sample can be obtained
from the MACS selection function and the local cluster X-ray
luminosity function. In a no-evolution scenario we expect to find 57
clusters at $z>0.3$ and $f_{\rm X,12} \ge 2$, and 151 at $z>0.4$ and
$f_{\rm X,12} \ge 1$. More than 20 MACS clusters are predicted to lie
at $z>0.6$.  This constitutes an improvement of a factor of about 50
over existing samples in the same redshift and luminosity range.

\section{Status and first results}
The MACS cluster sample currently comprises 41 systems with
spectroscopic redshifts in the range $0.3\le z\le 0.56$ (Fig.~2). An
additional 85 MACS sources with photometric redshifts $z\ge 0.3$ are
scheduled for spectroscopic observation.

The completeness of this preliminary sample cannot be easily
quantified.  However, the bright part of the MACS/BSC source list
($f_{\rm X,12} \ge 2$) has been mostly identified, and the
completeness of the sample of (presently) 25 clusters found above this
higher flux limit can be estimated. We thus use this subsample to
attempt to constrain cluster evolution at the very highest X-ray
luminosities ($L_{\rm X} > 1\times 10^{45}$ erg s$^{-1}$).  Even if we
knew nothing about the completeness of this subsample, the observed 25
clusters (where 57 are expected) already limit any negative evolution
to about a factor of two. If the estimated incompleteness of this
subsample is taken into account (R.A. range not yet fully covered,
softer X-ray sources not yet screened, etc), we expect to eventually
find 43 of the predicted 57 clusters. When the statistical and
systematic errors of measurement and prediction are taken into
account, the difference of about 15 clusters is statistically not
significant. Our tentative conclusion is that there is no significant
evolution of the X-ray cluster luminosity function out to $z\sim 0.4$
at any luminosity.\\*[-5mm]

\vfill

\begin{iapbib}{99}{
\bibitem{borgani} Borgani S. et al., {\it these proceedings},
                and references therein
\bibitem{bcs} Ebeling H. et al., 1998, MNRAS, 281, 799
\bibitem{nep} Gioia I. et al.,  {\it these proceedings},
                and references therein
\bibitem{warps} Jones L.R. et al., {\it these proceedings},
                and references therein
\bibitem{rdcs} Rosati P. et al., {\it these proceedings},
                and references therein
\bibitem{cfa} Vikhlinin A. et al., {\it these proceedings},
                and references therein

}
\end{iapbib}
\end{document}